\documentstyle[aps,multicol,epsfig]{revtex}

\begin{document}

\title{Operation characteristics of piezoelectric quartz tuning forks 
in high magnetic fields at liquid Helium temperatures}
\author{J.~Rychen, T.~Ihn, P.~Studerus, A.~Herrmann, and K.~Ensslin}
\address{Laboratory of Solid State Physics, ETH Z\"urich, 
CH-8093 Z\"urich, Switzerland}
\author{H.~J.~Hug, P.~J.~A.~van Schendel, and H.~J.~G\"untherodt}
\address{Institute of Physics, University of Basel, CH-4056 Basel, 
Switzerland}
\maketitle

\begin{abstract}
Piezoelectric quartz tuning forks are investigated
in view of their use as force sensors in dynamic mode scanning
probe microscopy at temperatures down to 1.5 K and in magnetic fields
up to 8 T.
The mechanical properties of the forks are extracted from the
frequency dependent admittance and simultaneous interferometric 
measurements. The performance of the forks in a cryogenic 
environment is investigated. Force-distance studies performed with
these sensors at low temperatures are presented.
\end{abstract}

\begin{multicols}{2}
\narrowtext

Piezoelectric quartz tuning forks have been introduced in scanning 
probe microscopy by G\"unther, Fischer and Dransfeld in Ref. 
\onlinecite{gunther} for use in scanning near field acoustic
microscopy and later by Karrai and Grober in Ref. \onlinecite{karrai}
as the distance control for a scanning near field optical microscope 
(SNOM).
Several other implementations of tuning forks have been 
reported, e.g. in SNOMs \cite{atia,ruiter,salvi,tsai}, scanning force 
microscopes (SFMs) \cite{edwards,giessibl}, magnetic force microscopes 
\cite{todorovic} and in the acoustic near field microscope \cite{steinke}.
The operation in a cryogenic environment was reported by 
Karrai and Grober in their pioneering work in Ref. \onlinecite{karrai}. 
To our knowledge 
operation characteristics at temperatures below 10K were not reported to 
date.
In this paper we present results on piezoelectric tuning 
fork sensors in our low temperature SFM which operates in the sample 
space of a $^4$He cryostat \cite{rychen}. 

In our studies we utilized commercially available 
tuning forks (see inset of Fig. 1a) which
are usually employed in watches with a standard frequency of $2^{15}$Hz. 
These forks are fabricated from wafers of
$\alpha$-quartz with the optical axis oriented approximately normal to the 
wafer plane.

The tuning fork can either be mechanically driven by an additional piezo 
element or electrically excited through the tuning fork 
electrodes \cite{karrai1}.
Similar to Ref. \cite{edwards} we drive the 
oscillation electrically by applying an 
AC-voltage of typically $U=0.01-10$mV
to the tuning fork contacts. For the investigation of the tuning fork
behavior we measure
the complex admittance of the fork with a two-channel lock-in
amplifier.
When employed as the sensor for dynamic force
measurements the tuning fork is
part of a phase-locked loop described in Ref. \onlinecite{rychen}.

\begin{figure}[b!]
\noindent\epsfig{file=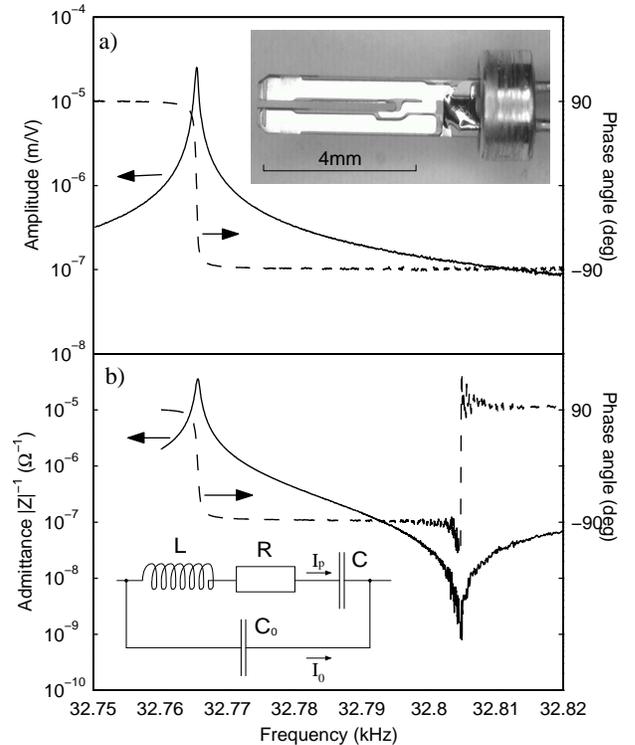,width=0.95\linewidth}
\caption{(a) Mechanical resonance measured at room 
temperature at a pressure of $6\times 10^{-7}$mbar with an optical 
interferometer. Inset: image of the tuning fork.
(b) Electrical tuning fork resonance measured simultaneously.
Inset: Equivalent circuit for the piezoelectric quartz 
tuning fork resonator. Solid and dashed lines are the respective amplitude 
and
phase.}
\label{fig1}
\end{figure}

Figure \ref{fig1}b shows a typical resonance in the admittance of a
tuning fork measured at room temperature at a pressure of 
$6\times 10^{-7}$ mbar. The admittance
exhibits an asymmetric resonance at 32768Hz and a sharp minimum at
about 30Hz above this resonance. The current through the fork
consists of two parts \cite{karrai1}: $I_{p}$ is the current created by the
mechanical (harmonic) oscillation of the fork arms through the 
piezoelectric
effect of the quartz. $I_{0}$ is the capacitive current through the
fork. The behavior of the admittance
can therefore be modeled with the equivalent circuit shown in
the inset of Fig. \ref{fig1}b. 

The LRC series resonator with a resonance frequency
$f_{0}=1/(2\pi\sqrt{LC})$ around $2^{15}$Hz and a quality factor
$Q=\sqrt{L/(CR^2)}$ which is typically of the order of $10^4$ allows
the current $I_{p}$ to pass. Using a mechanical model one can relate 
$L$, $R$ and $C$ with the effective mass of one arm $m$, the damping
constant $\gamma$, the spring constant $k$ and the driving force
$\alpha U$ via $L=m/\left(2\alpha^2\right)$, $C=2\alpha^2/k$,
$R=m\gamma/\left(2\alpha^2\right)$ \cite{rychen2}. The capacitance
$C_{0}$ is mainly determined by the geometrical arrangement of the
contacts on the crystal, the dielectric properties of the quartz and
by cable capacitances. The fit to the measured admittance in Fig. 
\ref{fig1}a
(which could not be distinguished in the plot from the measured curve)
leads to $C_{0}=1.2129$pF, $C=2.9fF$, $L=8.1\times 10^3$H, 
$R=27.1$k$\Omega$,
$f_{0}=32765.58$Hz and $Q=61730$.

In addition to the electrical resonance we measured the mechanical
resonance amplitude $x$ of one of the tuning fork arms (see Fig. 1a)
utilizing the interferometer setup usually used for optical
cantilever deflection detection in a scanning force microscope \cite{hug}. 
From a combination of both measurements (Fig. 1a and 1b) and using the 
relation
$I_{p}=4\pi f\alpha x$ \cite{karrai1} we determined the effective mass
$m=0.332$ mg, the quality factor $Q = 61734$, the spring constant
$k=14066.4$ N/m and the piezoelectric coupling constant 
$\alpha=4.26\mu$C/m. The effective mass calculated from the density of
quartz and the dimensions of a tuning fork arm according to Ref. 
\cite{karrai1}
turns out to be 0.36 mg, in good agreement with our measured value. 
A linear relation between the driving voltage and the oscillation
amplitude was found in the interferometer measurement down to amplitudes
of 1 nm as well as in large-amplitude measurements performed under an
optical microscope up to amplitudes of about 100 $\mu$m.

For the use in our SFM we remove the tuning forks from
their casing and glue a thin metallic wire (diam. 10$\mu$m - 50$\mu$m)
in the direction of the oscillatory 
motion to the end of one prong. The wire is then
etched electrochemically to form a sharp tip. If the wire is electrically
connected to one of the tuning fork contacts its length is about 500$\mu$m.
In cases where we connect the wire to a separate contact pad the wire can
be up to 3mm long. The additional weight $\Delta m$ fixed to the
tuning fork arm is in the range between 1.5$\mu$g and
50$\mu$g. In
order to obtain the most sensitive force gradient detection it is
important to keep the relative mass increase as small as possible.
After this modification the resonance of the tuning forks are always
shifted to lower frequency, in most cases less than 100Hz and typical
quality factors $Q=10000$ under ambient conditions are reached.

In our SFM we operate the tuning forks in the 
gas flow of a variable temperature $^4$He cryostat. Alternatively the
sample space can be flooded with liquid He and the microscope is then
operated either in normal fluid $^4$He or in a mixed
normal-superfluid phase (below 2.2K) \cite{karrai2}. 
Table \ref{table1} shows tuning
fork resonance characteristics obtained under these different conditions.
Compared to operation in the gas the resonance frequency of the fork
is shifted by more than 500Hz to lower frequencies in the normalfluid 
liquid.
At the same time the $Q$-value of the resonator decreases due to the 
significantly increased friction in the liquid. At temperatures below 
2.2K the quality factor increases again as a result of the formation of
superfluid $^4$He which tends to suppress friction effects \cite{karrai2}. The 
explanation of the reduction of the resonance frequency in the superfluid
is an open question. These measurements demonstrate the robustness of
the tuning fork properties when external conditions are severly
changed. With a conventional cantilever for scanning force microscopy
it is difficult to achieve $Q$-values of this order in liquid
$^4$He. However, even with the tuning forks, operation in liquid He is
cumbersome because the resonance frequency fluctuates strongly. During
scanning the tuning fork is typically operated at a constant frequency
shift of 100 mHz and deviations of more than a few mHz induced by the
environment are intolerable. 

\begin{figure}
\noindent\centering\epsfig{file=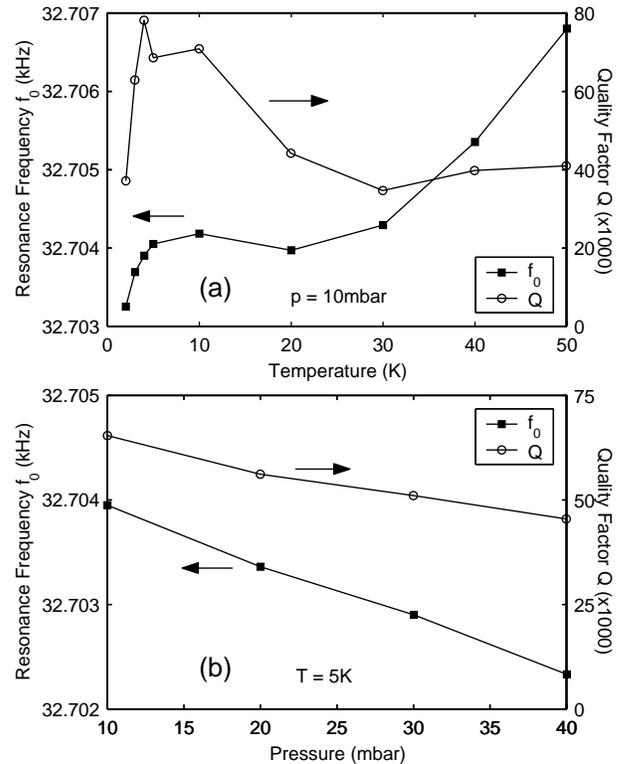, width=0.95\linewidth}
\caption{Measured resonance frequency and $Q$-value (a) versus
temperature at constant pressure of 10mbar and (b) versus pressure at
a constant temperature of 5K.}
\label{fig2}
\end{figure}

The temperature coefficient of the resonance frequency below 5K was
determined to be $260\mathrm{mHz/K}$ at a constant pressure of
10 mbar (see Fig. \ref{fig2}). The pressure coefficient was $50\mathrm{mHz/mbar}$ at 5 K. This means that frequency shifts of the order of 10mHz 
are produced by temperature instabilities of about 50mK or pressure 
instabilities of about 0.2mbar. In addition, we measured the dependence of 
the resonance frequency on an external magnetic field in
the range between 0 and 8 T. The detected frequency shift was smaller
than 100mHz. 

\begin{table}[b]
\caption{The resonance characteristics of the tuning fork in different 
helium phases\label{table1}}
\begin{tabular}{lrrr} 
 & $T$ (K) & $f_0 \mathrm{(Hz)}$ & $Q$ \\
\tableline
helium gas ($p<1$ mbar)& $\approx 5$ & 32634 & 22665 \\
liquid helium & 4.2 & 32110 & 2153 \\
superfluid helium & 1.56 & 32056 & 7583 \\
\end{tabular} 
\end{table}

\begin{figure}
\noindent \centering \epsfig{file=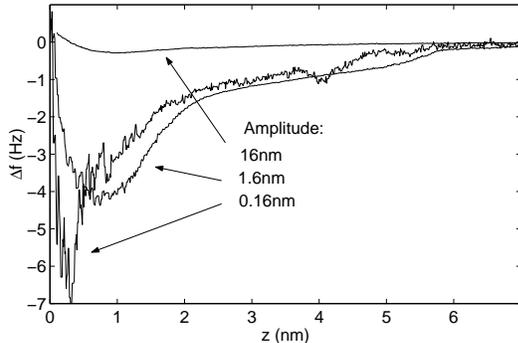, width=0.8\linewidth}
%\pdfimage width 0.8\linewidth figure3.png
\caption{Measured frequency shift as a function of tip sample 
separation for different oscillation amplitudes ($T$=2.5K).}
\label{fig4}
\end{figure}

In order to demonstrate the power of the piezoelectric tuning fork 
sensing of tip-sample interactions we show in Fig. \ref{fig4} a set of
measuements of the frequency shift versus distance of the tip to the
Au surface measured at 2.5K and zero tip-sample voltage. From the
quality of topographical images 
taken with this tip just before these measurements we deduce a tip 
radius of several hundred nanometers. 

The oscillaton amplitude was varied by a factor of 10 from one curve 
to the next, starting from a sub-nm value. A larger oscillation 
amplitude averages over a larger $z$-range in the repulsive as well 
as in the attractive region of the interaction such that the frequency 
shift at a given distance decreases with increasing amplitude. Such a 
behavior was quantitatively described by Giessibl \cite{giessibl2} for 
an attractive van der Waals-potential of the form $V\propto z^{-n}$, 
where $n$ is a positive integer.

In the following we discuss several aspects of the utilization of tuning
forks in SFMs at cryogenic temperatures.
The typical values of the spring constants of tuning forks are much higher
than the ones of conventional SFM cantilevers. This has several 
implications
for their use in a dynamic mode SFM: first, a given force gradient leads to
a smaller shift of the resonance frequency than in conventional 
cantilevers,
since typically $\delta f/f_{0}\propto k^{-1}$, no matter whether
small or large oscillation amplitudes are used \cite{giessibl2}. Care
has to be taken in the design of the control electronics to make sure
that frequency shifts of at least 10mHz can be measured to compensate
this disadvantage. Second, there is no danger for the tip to snap into
contact with the sample, since the condition $k>\partial F/\partial z$
is met for all tip-sample spacings. This makes the tuning fork an
ideal tool to investigate tip-sample interactions as a function of
distance. And last but not least, the high spring constant makes
tuning forks ideal for specific applications, e.g. for nanolithography
in the non-contact mode \cite{held} or as carriers for all kinds of
scanning nano-sensors which may be harder to implement on conventional
SFM cantilevers. These issues will be discussed in future publications.

The $Q$-values obtained with our tuning forks at pressures around
1mbar are generally of the same
order as the best cantilevers when operated under UHV-conditions. The
robustness of $Q$ against pressure changes is also significantly higher
than that of conventional cantilevers. Compared to piezoresistive 
cantilevers which tend to heat systems at low temperatures with 
powers in the 1mW range tuning forks do not produce any significant 
amount of heating power and are therefore ideal for future 
applications in $^3$He-systems or dilution refrigerators.

In conlusion, we have demonstrated the operation of piezoelectric 
tuning forks as sensors for dynamic mode scanning force microscopy
at cryogenic temperatures and discussed their performance. The 
robustness of this sensor allows to achieve very high quality factors 
even under the otherwise problematic conditions of non-UHV environments. 
The force gradient detection method is well suited for force distance 
studies.

Financial support by ETH Z\"urich is gratefully acknowledged.

\end{multicols}
\end{document}